\documentclass[twocolumn,nopacs,floatfix,superscriptaddress]{revtex4-2}

\usepackage{amsmath}
\usepackage{graphicx}
\usepackage{epsfig}
\usepackage{color}
\usepackage{soul}
\usepackage{longtable}
\usepackage{verbatim}
\usepackage[OT1]{fontenc}
\usepackage[normalem]{ulem}
\usepackage{hyperref}
\hypersetup{colorlinks=true, citecolor=blue, urlcolor=blue, linkcolor=blue}
\def\CC{C^{(c)}}

\def\GS#1{S^W_{#1}}

\begin{document}
 
\title{Scale-free correlations in the dynamics of a small ($N \sim 10000$) cortical network}

\author{Sabrina Camargo}
\email{scamargo@unsam.edu.ar}
\affiliation{Instituto de Ciencias F\'isicas (ICIFI-CONICET), Center for Complex Systems and Brain Sciences (CEMSC3), Escuela de Ciencia y Tecnología, Universidad Nacional de Gral. San Martín, Campus Miguelete, 25 de Mayo y Francia,  1650, San Martín, Buenos Aires, Argentina}
\affiliation{Consejo Nacional de Investigaciones Cient\'{\i}fcas y Tecnol\'ogicas (CONICET), Godoy Cruz 2290, 1425, Buenos Aires, Argentina.}

\author{Daniel A. Martin}
\affiliation{Instituto de Ciencias F\'isicas (ICIFI-CONICET), Center for Complex Systems and Brain Sciences (CEMSC3), Escuela de Ciencia y Tecnología, Universidad Nacional de Gral. San Martín, Campus Miguelete, 25 de Mayo y Francia,  1650, San Martín, Buenos Aires, Argentina}
\affiliation{Consejo Nacional de Investigaciones Cient\'{\i}fcas y Tecnol\'ogicas (CONICET), Godoy Cruz 2290, (1425), Buenos Aires, Argentina.}

\author{Eyisto J. Aguilar Trejo}
\affiliation{Instituto de Ciencias F\'isicas (ICIFI-CONICET), Center for Complex Systems and Brain Sciences (CEMSC3), Escuela de Ciencia y Tecnología, Universidad Nacional de Gral. San Martín, Campus Miguelete, 25 de Mayo y Francia,  1650, San Martín, Buenos Aires, Argentina}
\affiliation{Consejo Nacional de Investigaciones Cient\'{\i}fcas y Tecnol\'ogicas (CONICET), Godoy Cruz 2290, (1425), Buenos Aires, Argentina.}

\author{Aylen de Florian}
\affiliation{Instituto de Ciencias F\'isicas (ICIFI-CONICET), Center for Complex Systems and Brain Sciences (CEMSC3), Escuela de Ciencia y Tecnología, Universidad Nacional de Gral. San Martín, Campus Miguelete, 25 de Mayo y Francia,  1650, San Martín, Buenos Aires, Argentina}

\author{Maciej A.\ Nowak}
\affiliation{Mark Kac Center for Complex Systems Research and Institute for Theoretical Physics, Jagiellonian University, Krak\'ow, Poland}

\author{Sergio A.\ Cannas}
\affiliation{Facultad de Matem\'atica Astronom\'ia F\'isica y Computaci\'on, Universidad Nacional de C\'ordoba, Instituto de F\'isica Enrique Gaviola (IFEG-CONICET), Ciudad Universitaria, 5000 C\'ordoba, Argentina.}
\affiliation{Consejo Nacional de Investigaciones Cient\'{\i}fcas y Tecnol\'ogicas (CONICET), Godoy Cruz 2290, (1425), Buenos Aires, Argentina.}

\author{Tom\'as S.\ Grigera}
\affiliation{Instituto de F\'isica de L\'iquidos y Sistemas Biol\'ogicos (IFLySiB), CONICET and Universidad Nacional de La Plata, Calle 59 n 789, 1900 La Plata, Argentina}
\affiliation{Departamento de F\'isica, Facultad de Ciencias Exactas,  Universidad Nacional de La Plata, 1900 La Plata, Argentina}
\affiliation{Consejo Nacional de Investigaciones Cient\'{\i}fcas y Tecnol\'ogicas (CONICET), Godoy Cruz 2290, (1425), Buenos Aires, Argentina.}
\affiliation{Istituto dei Sistemi Complessi, Consiglio Nazionale delle Ricerche, via dei Taurini 19, 00185 Rome, Italy}

\author{Dante R.\ Chialvo}
\affiliation{Instituto de Ciencias F\'isicas (ICIFI-CONICET), Center for Complex Systems and Brain Sciences (CEMSC3), Escuela de Ciencia y Tecnología, Universidad Nacional de Gral. San Martín, Campus Miguelete, 25 de Mayo y Francia,  1650, San Martín, Buenos Aires, Argentina}
\affiliation{Consejo Nacional de Investigaciones Cient\'{\i}fcas y Tecnol\'ogicas (CONICET), Godoy Cruz 2290, (1425), Buenos Aires, Argentina.}

\date{August 8, 2023}

\begin{abstract}
The advent of novel opto-genetics technology allows the recording of brain activity with a resolution never seen before.  The characterisation of these very large data sets offers new challenges as well as unique theory-testing opportunities.  Here we discuss whether the spatial and temporal correlation of the collective activity of thousands of neurons are tangled as predicted by the theory of critical phenomena. The analysis shows that both, the correlation length $\xi$ and  the correlation time $\tau$ scale as predicted as a function of the system size.  With some peculiarities that we discuss, the analysis uncovers new evidence consistent with the view that the large scale brain cortical dynamics corresponds to critical phenomena.
\end{abstract}

\maketitle

\section{Introduction}
\label{sec:introduction}

The study of correlation functions is central to understanding critical phenomena throughout disciplines 
\cite{cavagna2010, Attanasi2014, cavagna_physics_2018, grigera2021}. The correlations of an infinitely large system poised near a critical point, will vanish at infinity as a power-law, i.e.\ rather slowly.  In other words, the entire system seems to be correlated.  Conversely, away from criticality, correlations decay exponentially fast following closely the (typically short range) interactions.  If the system is critical but not infinite, the power law is altered by the finite system size, but there is a characteristic dependence of the correlations on system size, at criticality, which can be exploited \cite{Attanasi2014} to establish whether the system exhibits critical correlations.  This finite-size behaviour has also been used as a proxy to determine if the brain exhibits critical dynamics, including attempts at very large scale \cite{fraiman_what_2012, expert2011, haimovici_brain_2013} or on a relatively sub-sampled regime \cite{ribeiro2020, ribeiro2021}.

Time correlations, although less studied in the biological case, also have a characteristic behaviour at criticality, known as dynamic scaling.  Dynamic scaling means that space and time correlations are intertwined.  A basic statement of dynamic scaling is that the correlation \emph{temporal scale} (of a collective quantity) grows as a power law of the correlation \emph{spatial scale.}

In these notes we make an exhaustive exploration of correlations, both spatial and temporal, as a function of size, analysing a  large collection of neurons  recorded from mice visual cortex using opto-genetic techniques.  Our analysis computes  correlations inside boxes of increasing size $W$, based on our recent demonstration that this approach is equivalent to changing the system size $L$ \cite{martin2021}. The robustness of spatial correlations were tested by using three different methods: two for the connected correlation function and the density correlation function.  We similarly study time correlations in boxes of different sizes, as well as the relationship between characteristic correlation lengths and times.

The paper is organized as follows: First the data is described. After that, a subset of data is used to introduce the correlation methods.  Next, we describe the main results starting with the finite-size dependence of the spatial correlations  and followed by the temporal correlations. The two results are combined to asses the presence of dynamic scaling, i.e., the dependence of temporal fluctuations on the correlation length. Finally, the correlation matrix is analyzed in terms of the scale-invariance of its eigenvalues spectra. The paper closes with a discussion of the caveats and limitations as well as some future work.

\section{Methods}
\label{sec:methods}

\subsection{Experimental data} The data analyzed here is freely available \cite{stringer2018} and fully described by Stringer \textsl{et al.} \cite{stringer2019}.  These authors have provided the activity time series of a set of approximately 10,000 neurons in the visual cortex of several mice.  Animals were awake, head-fixed but able to run freely over an air-floating ball, and had been implanted with 3 to 4$\,$mm cranial windows centered over their visual cortex.  Data were recorded in eleven planes of a region of $\sim 1\,$mm$^{2}$ of the visual cortex, while the mouse is not receiving  any particular visual stimulation (it is watching a dark screen).  Neuronal activity was obtained optically, and the data provided includes the position $(x,y,z)$ of each neuron, and its activity sampled at a rate of $2.5$ or $3\,$Hz during 21055 frames. Recordings were performed using multiplane acquisition controlled by a resonance scanner, with 11 planes spaced 35 $\mu m$ apart in depth.   The activity data analysed here corresponds to the de-convolution of the raw neuronal Ca$^{2+}$ signal (called ``Fsp signal'' in ref.~\cite{stringer2019}), hereafter denoted as $S(t)$, giving spike counts per time bin.  These data were obtained by the authors of ref.~\cite{stringer2019} from the Ca$^{2+}$ images using Suite2P \cite{Suite2P}, which performs motion correction, cell detection and neuropil correction.  After that, the OASIS spike deconvolution algorithm was applied \cite{RobustWithOASIS, OASIS}.  Full details on the imaging methods and animal protocols can be found in ref.~\cite{stringer2019}.

We analyze correlations in neuronal activity both in the $S(t)$ signal and in a derived  point process (PP), constructed by thresholding the $S(t)$ time series (Eq.~\eqref{eq:point-process} below).  Fig.~\ref{fig:data} shows a schematic of the animal setup (panel A) and examples of the signal $S(t)$ (panel B) and the derived point processes (panel D).  Illustrative examples of the distribution of $S(t)$ are given in panel C.  The analysis is done on nine data-sets from seven mice, which for simplicity we  labeled consecutively  \cite{label1}.  The field of view in each of the eleven planes spans a range of $x\in[4, 1010]\,\mu$m and $y\in[4,1012]\,\mu$m.  The neurons recorded are located, from the more superficial to the deeper ones,  at planes with coordinates $z=70, 105, 140, 175, 210, 245, 280, 315, 350, 385, 420\, \mu$m, for planes 1 to 11 respectively.

\begin{figure}[ht!]
    \centering
    \includegraphics[width=0.95\columnwidth]{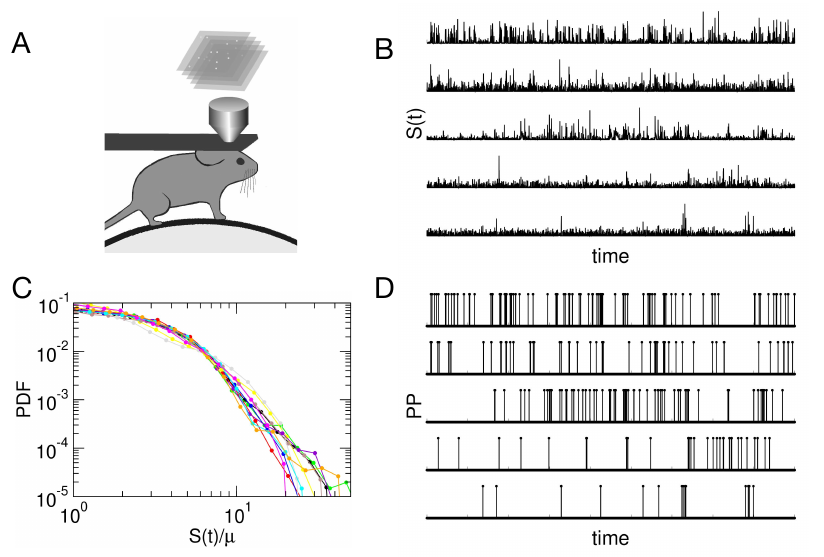}
    \caption{Schematics of the experimental data.  (A) Sketch of the experimental setup: a head-fixed mouse is able to run on a spherical treadmill while the brain activity is monitored via a multiplane Ca$^{2+}$ opto-genetic imaging.  (B) Examples of the $S(t)$ signal time series.(C) Typical probability distribution of the $S(t)$ signal time series for nine neurons, after normalization by its mean ($\mu$). (D)  Examples of point process extracted from the $S(t)$ signal. }
    \label{fig:data}
\end{figure}

\subsection{Correlation analysis} The exploration of the correlation properties is aimed at establishing up to what extent the spatial and temporal correlations are entangled, as in other systems exhibiting critical dynamics. For that purpose suitable correlation functions \cite{grigera2021} in space and time  need to be defined and computed. 

\paragraph{Spatial connected correlation function---} At the risk of being redundant, let us remark that the connected correlation function (CCF) has different properties than the (more usual) Pearson correlation function computed between two or more variables. Formally, the (space-averaged) CCF is
\begin{equation}
C(r)=\dfrac{1}{c_0}\dfrac{\sum_{i,j}u_iu_j\delta(r-r_{ij})}{\sum_{i,j}\delta(r-r_{ij}) }
\label{eq:connected_corr}
\end{equation}
where $r_{ij}$ is the Euclidean distance between the given pair of neurons, $\delta$ is a Dirac delta selecting the pairs separated by a distance $r$,  $u_i(t) = S_i(t) - \bar{s}(t)$,  and $\bar{s}(t)=\frac{1}{N}\sum_i^N S_i(t)$, and $S_i(t)$ is the activity of neuron $i$ at time $t$. Notice that the mean $\bar{s}(t)$ (often, in this context, called population mean), is the instantaneous spatial mean, and is subtracted from each signal at each time step. In that way, any confound common to the two (or more) signals is canceled out. For instance, an external drive to the entire neuronal population under study can increase Pearson correlations for all neuron pairs, but won't be affecting the value of $C(r)$ in Eq.~\ref{eq:connected_corr} (an illustrative example is shown in Fig. \ref{Arousal} in the Appendices).   Thus, $C(r)$ in Eq.~\ref{eq:connected_corr} describes the decay of correlations in space between the remaining \emph{fluctuations around the mean}, often called residual correlations (Ref. \cite{grigera2021} discusses the properties of the CCF at length, as well as the algorithms to compute it).

\paragraph{Spatial point process---} The time series $S(t)$ is extracted via de-convolution of the Ca$^{2+}$ fluorescence recordings. It represents the spike count per time bin (i.e., $ S(t)\geq 0$) being recorded at that particular area of the field of view, within a given sampling interval. For completeness, here we also consider a transformation of $S(t)$ into a point process (also called point field). The idea is to determine  possible effects of different signal-to-noise ratios by selecting only the most significant neuronal events to compute correlations.  In short, we define
\begin{equation}
  P_i(t) = \begin{cases} 0 & \text{if $S_i(t)=0$}, \\ 1 & \text{if $S_i(t)>0$},
  \end{cases} \label{eq:point-process}
\end{equation}
i.e.\ $P(t)$ is 1 if the neuron has fired during the observation window, and 0 otherwise.  We compute the CCF using the same definition Eq.~\eqref{eq:connected_corr} with $P(t)$ in place of $S(t)$, which we denote by $C_p(r)$.

\paragraph{Spatial counting statistics---} In addition to the previous two ways of measuring correlations, we consider a counting statistic approach.  We calculate at each time step $t$ the density $G(r)$ of spiking neurons (i.e., points) falling inside a thin shell of radius $r$ centered on a spiking neuron. The definition of G(r) is closely related to the radial distribution function commonly used to characterize the structure of liquids and solids. After proper normalization, this counting statistic is equivalent to $C_p(r)$ since both are defined for the same point process.  $G(r)$ is computed according to
\begin{equation}
  G(r)= \overline{ \dfrac{\rho N}{\rho_a(t) N_a(t)}
        \sum_{i,j} \dfrac{P_i(t) P_j(t) \delta(r-r_{ij}) }{\delta(r-r_{ij})} },
\label{eq:correlation_density}
\end{equation}
where $N$, $\rho$ are respectively the total number and average density of neurons, and $N_a(t)$, $\rho_a(t)$ the same quantities for \emph{active} neurons at time $t$, i.e.\ such that $P_i(t)=1$.  The overline indicates average over all time frames.  In other words, $G(r)$ is the ratio of active pairs over total pairs at distance $r$, normalised to obtain 1 when both populations correspond to uncorrelated Poisson processes.

\paragraph{Box scaling and correlation length---}  The correlation length is an indicative measure of the spatial extent of correlations.  There are several possible procedures to obtain in practice a correlation length from an experimental space correlation function, but the important point is that when trying to establish whether correlations are scale-free one needs to study the dependence of the experimental correlation scale with system size, or with observation scale \cite{martin2021}.  In our case it is clearly impossible to consider systems of different size, so we use the box-scaling procedure, measuring the correlation functions within a spatial observation window, or box, of linear size $W$.  In practice, we computed Eq. \eqref{eq:connected_corr}, considering only neurons within a square box of linear size $W$, and time averaged over all time frames to decrease statistical error:  

\begin{equation}
C_\mu^W(r)=\dfrac{1}{c_{0\mu}}\dfrac{\sum_{i,j \in W_\mu} \overline{u_iu_j}\delta(r-r_{ij})}{\sum_{i,j  \in W_\mu}\delta(r-r_{ij}) },
\label{eq:connected_corrU}
\end{equation}
  where the sum includes all pair of neurons that belong to the $\mu$-th box of the space grid of size $W$, and $u_i$ is the fluctuation of neuron $i$ signal with respect to the \emph{instantaneous} spatial average activity in the box $W_\mu$. After that, we averaged over as many non-overlapping boxes as possible. A similar procedure is performed for the spatial counting statistics, Eq. \eqref{eq:correlation_density}. 
We use the notation $C(r,W)$, $C_p(r,W)$, $G(r,W)$ to indicate the correlation functions restricted to a box.  In this case it is convenient to define, for the $S(t)$ signal, a length $\xi_0(W)$ such that $C(\xi_0,W)=0$, because $\xi_0(W)$ will grow linearly with $W$ if the system is scale free \cite{grigera2021, cavagna_physics_2018, martin2021}.  Similarly, for the $P(t)$ signal we use $C_p(\xi_{0p},W)=0$ and $G(\xi_{0G},W)=1$ for the spatial counting case.

\begin{figure*}
    \centering
    \includegraphics[width=1\textwidth]{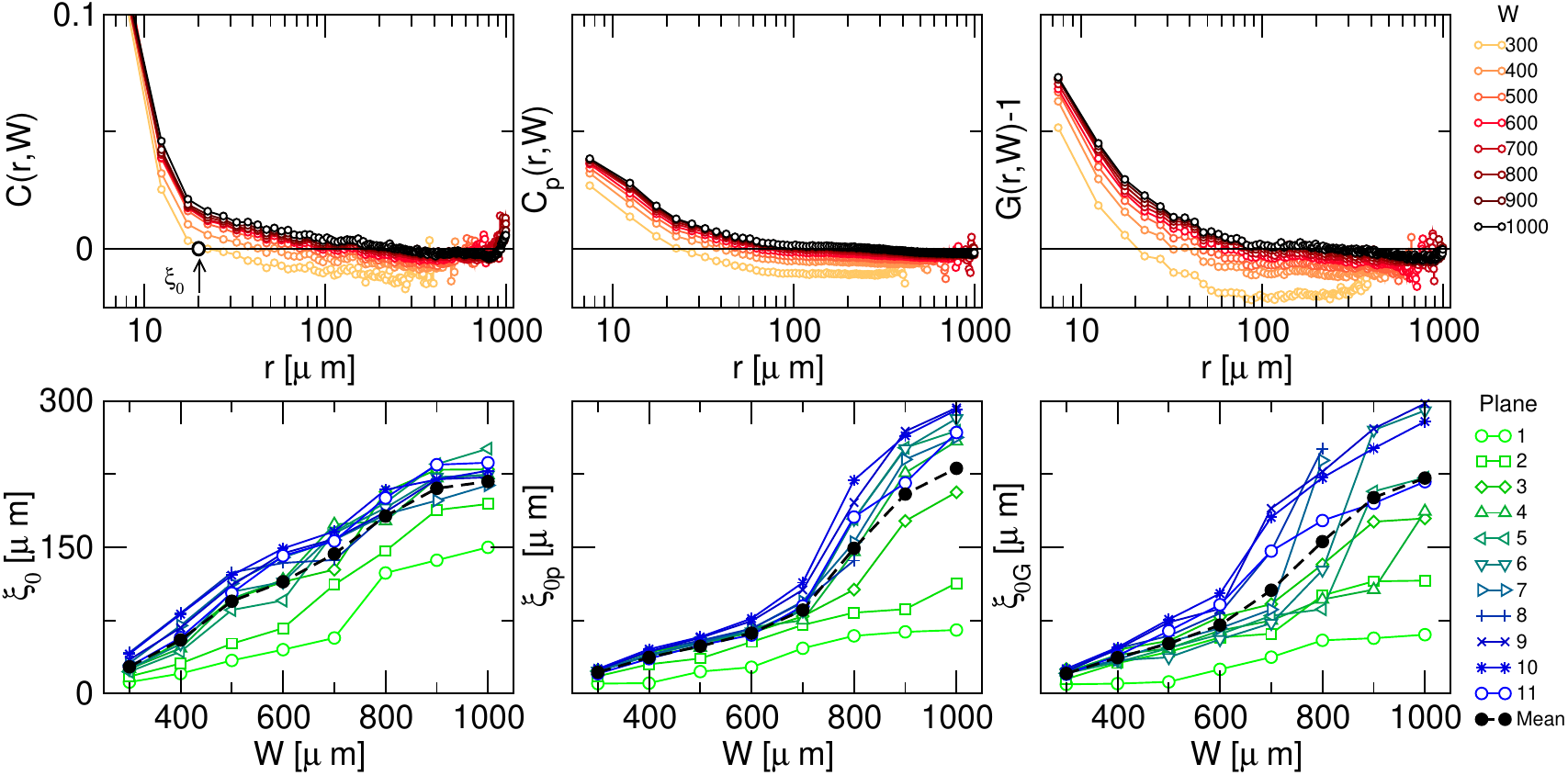}
    \caption{Examples of correlation and density functions (top panels) as a function of distance for plane 4 of mouse4c and different box sizes.  The bottom panels show the correlation length extracted for all planes from the three approaches: connected correlation function $CCF_r$ (left),  connected correlation function computed from the point process $CCF_p$ (center), and density function $G(r)$ (right).  Results are an average over 9 rotations of the box grid relative to the field of view.}
    \label{fig:methods}
\end{figure*}

Fig.~\ref{fig:methods} shows examples of the three approaches (for mouse 4c).
The top row illustrates the functions (for a single plane) from which the $\xi_0$ are extracted. The top left panel corresponds to the connected correlation function vs.\ distance $r$ for different box sizes computed from the $S(t)$ signals, where the arrow denotes $\xi_0$ for $W=300$, as an example. The top centre panel shows the connected correlation function for the point process, and the top right panel the results for the density function $G(r)$.  Results from all planes in this mouse are condensed in the bottom row.  Shown are the scaling of the correlation length $\xi_0$ with box size $W$ for each of the eleven planes using each of the three approaches. To prevent a possible bias given by the inhomogeneous distribution of the neurons' spatial locations the boxing of the sample was performed over 9 different rotations of the box grid relative to the field of view. 

\paragraph{Time correlations---} Time correlation functions are a measure of how correlations in a time series decrease as one compares two signals measured at increasing time intervals.  To assess how time and space correlations are intertwined, one studies the time correlations of spatially extended quantities.
We define
\begin{equation}
  \GS{\mu}(t)= \sum_{i\in W_\mu} S_i(t),
\end{equation}
where the sum includes all neurons that belong to the $\mu$-th box of the space grid of size $W$.  The connected time correlation is then
\begin{equation}
  \CC(t,W) = \frac{1}{N_W}\sum_\mu \frac{1}{T-t} \sum_{t'=0}^{T-t} \delta \GS{\mu}(t') \delta \GS{\mu}(t'+t), 
                       \label{eq:conn-corr-estimate}
\end{equation}
where $\delta \GS{\mu}(t) = \GS{\mu}(t) - (1/T)\sum_{t'=0}^T \GS{\mu}(t')$ and $N_W$ is the number of boxes of side $W$.

From the decay of each time correlation function one can extract a characteristic time scale, or correlation time $\tau$.    Rather than using a threshold, we found the spectral relaxation time  of Eq. \eqref{eq:conn-corr-estimate} is less prone to noisy fluctuations .  It is given as the solution of
\begin{equation}
  \label{eq:HHrelaxtime}
   \int_0^\infty \!\! \frac{dt}{t} \,\frac{ \CC(t)}{\CC(t=0)}
   \sin\left(\frac{t}{\tau}\right) = \frac{\pi}{4},
\end{equation}
(see  Appendix A  and  \cite{halperin_scaling_1969}  for the rationale behind this definition).

\begin{figure}
    \centering
    \includegraphics[width=0.81\columnwidth]{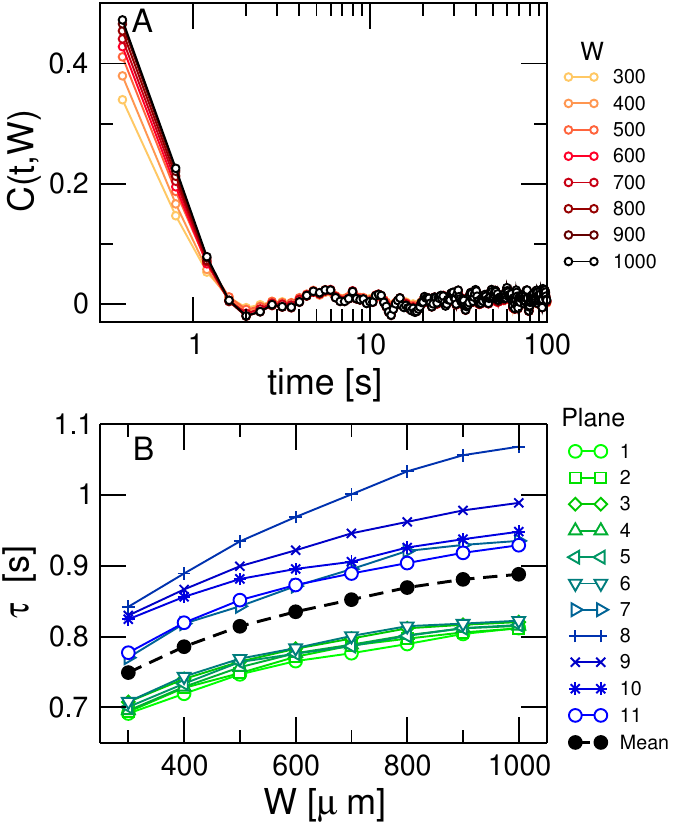}
    \caption{Correlation time analysis. (A) Examples of the time correlation functions computed from  the activity in on plane as a function of box size $W$. (B)  Correlation time $\tau$ as a function of box size $W$ for all planes (data from mouse 4c, the functions in panel A correspond to plane 5).}
    \label{fig:tau_example}
\end{figure}

An example of the correlation functions $\CC(t,W)$ and associated correlation times for one mouse is shown in Fig.~\ref{fig:tau_example}.

\paragraph{Computer code---} The code  used for the computation of the quantities defined above can be obtained from GitHub \cite{github} or Zenodo \cite{zenodo}.

\section{Results}
\label{sec:results}

\paragraph{Spatial correlations---} From the three measures of correlation it follows that the more correlated a pair of neurons is, the closer in space the members of the pair tend to be.  This observation is not entirely trivial, since the fact that neurons can develop very long axons (up to several hundred times the size of the neuron' soma) makes it possible that the interaction develops with many non-local contacts in a way that allows direct transmission of information to extremely far away neurons, making the Euclidean distance irrelevant.  On the other hand, it is clear that developing a longer axon carries a larger energy cost, so that the Euclidean distance should play a role, even if indirect, after all.  In fact the three correlations $C(r)$, $C_p(r)$ and $G(r)$ show clearly that this is the case.  Additional support for distance decay of correlations is obtained through the reverse procedure of picking the pairs within a range of a given value of correlation and computing their average distance.  The results of these computations are fully consistent with the correlations already commented  (see examples of these calculations in the Appendices).  Thus, each of the strategies used here confirm that there is a distance dependence of the correlations.  This is in contrast with the interpretation of the Pearson pair correlation results for the same data given in ref.~\cite{stringer2019}.

\begin{figure}
    \centering
    \includegraphics[width=\columnwidth]{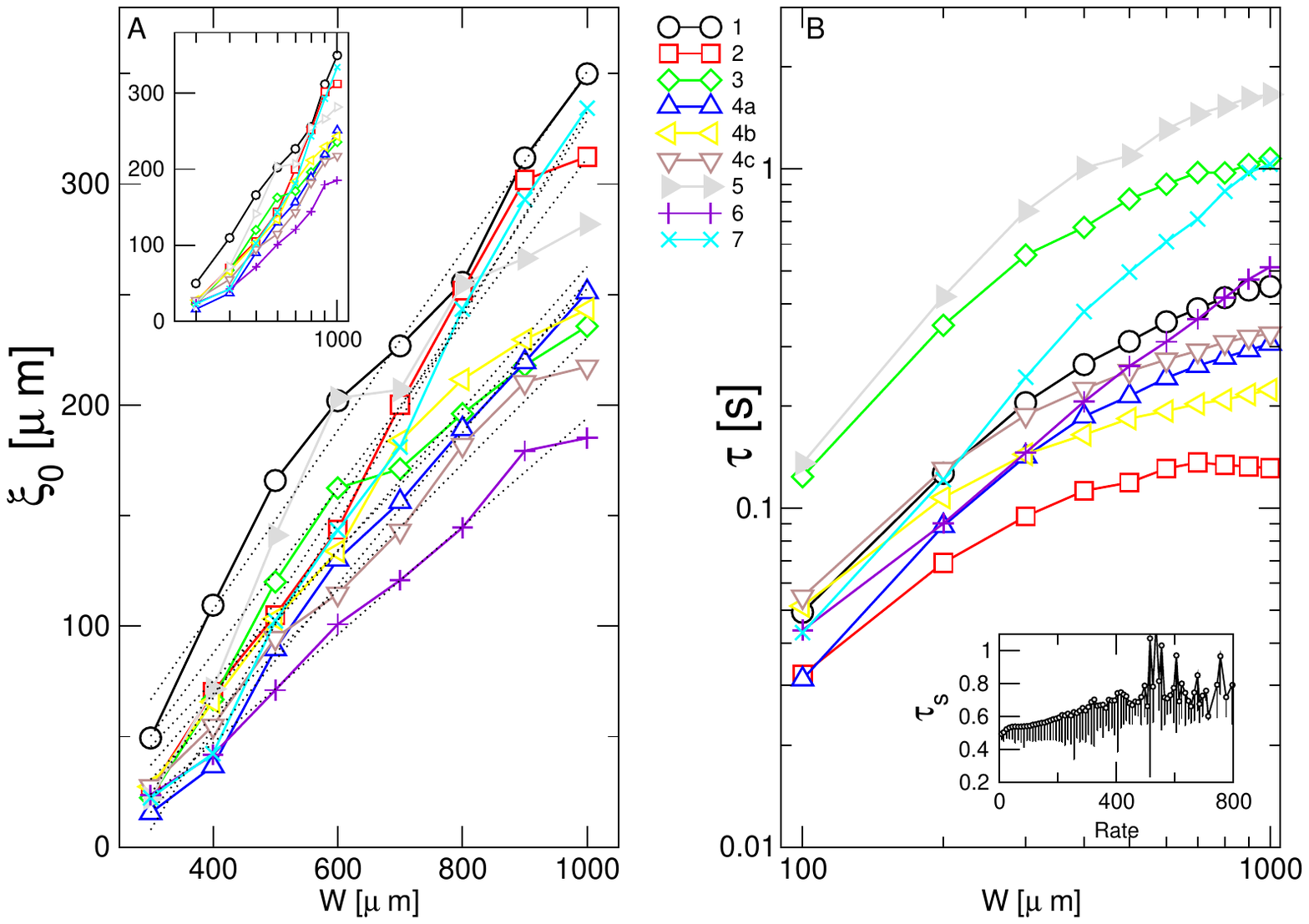}
    \caption{Size-dependence of correlation length and correlation time. (A) Characteristic correlation length $\xi_0$ vs.\ box size $W$, averaged over all planes for each mouse.  Windows smaller than $W\approx 300\,\mu$m yield values of $\xi_0$ less than the soma size and are therefore excluded.  The curves are compatible with a linear growth of $\xi_0$ with $W$.  This is evidence that the correlations of neuronal activity fluctuations are scale-free.  (A, inset) Same data in a semi-logarithmic plot: note the steeper increase of $\xi_0$ for larger $W$ indicating that their relation is not logarithmic. (B) Correlation time $\tau$ for the $\GS{}(t)$ vs box size $W$ for each mouse.  Correlation functions were averaged over all boxes in the grid and over 9 rotations of the grid relative to the field of view, then the correlation time is averaged over all planes. (B, inset) Correlation time   of a single neuron $\tau_s$  versus its own mean activity (spikes per second).  Plotted is $\tau_s$ averaged in the corresponding activity bin of width 10$\,s^{-1}$,  using all neurons from all mice and planes.}
    \label{fig:r0_ave_all}
\end{figure}

We proceed now to measure the spatial scale of the correlation decay, i.e.\ the correlation length (from now on we focus on $C(r)$ since the other correlation functions yield similar results).  We compute $C(r;W)$ and $\xi_0(W)$, given by $C(\xi_0(W);W)=0$, on boxes of side $W$ ranging from $300\,\mu$m to $1\,$mm for each plane of each mouse ($\xi_0$ vs.\ $W$ averaged over all planes is shown for all mice in Fig.~\ref{fig:r0_ave_all}).  We find that $\xi_0$ grows linearly with $W$: this observation is crucial, because it implies that the system is \emph{scale free} \cite{cavagna_physics_2018, grigera2021}, i.e.\ that the correlation length is larger than the system size, with the consequence that the scale for decorrelation is given by the system size, or by the observation window $W$.  If there was a correlation scale smaller than the system size one would have \emph{logarithmic}, rather than linear, growth of $\xi_0$, which is not what we observe (Fig.~\ref{fig:r0_ave_all}, inset).

At this point, we should consider potential artifacts, such as  head motion  and changing levels of animal arousal which are known to confound the calculation of the usual pair-wise Pearson correlations. In contrast, we are able to demonstrate that the type of correlations reported in this work are completely inmune to these artifacts. The reason is related to the fact that the connected correlation function used here is computed from the \emph{fluctuations around the instantaneous spatial average of activity} (Eq.~\eqref{eq:connected_corr}), thus eliminating common drives produced, for example  by motion-induced optical artifacts or by slow trends in excitability or by behavioral changes.  We  illustrate the robustness of these calculations, in Appendices Sec.~\ref{sec:artifacts} by conducting a series of numerical simulations, using the neuronal network model of \cite{Eyisto} (a network of cellular automaton neurons defined in \cite{haimovici_brain_2013}, running on a 2D lattice). There we show that signal noise (i.e., mimicking motion), time binning, or changing levels of arousal, do not change significantly the behavior of the scaling of the correlation length with window size, providing confidence on the robustness of the present results.

\paragraph{Time correlations---} In physical systems near criticality, the dynamical behaviour displays specific characteristics alongside the scale-free properties of the static correlations.  To asses to what extent the phenomenology of neuronal systems can be described with a formalism similar to that of equilibrium critical systems, we study time correlations of single neurons and of the collective signal $\GS{}(t)$ in the same boxes we used for the space correlations.  For each mouse and plane we computed $\CC(W,t)$ and extracted a correlation time as described in Methods.
We have observed that the correlation time of a single neuron,  $\tau_s$, which is the relaxation time computed from Eq.\eqref{eq:conn-corr-estimate} using single neuron timeseries,  grows with the neuron's activity (i.e., firing rate per unit time), see Fig.~\ref{fig:r0_ave_all}, inset of panel B.  This effect is an artifact of the deconvolution procedure.  Since the aim of the computation in boxes is to gauge how $\tau$ is affected by the collective behaviour of the interacting neurons, we have subtracted    $\tau_s$   from the values of correlation time obtained from the box signal (this was done plane by plane).  All correlation times reported are subject to this subtraction.

The correlation time (averaged over planes) is shown in Fig.~\ref{fig:r0_ave_all}B.  Since we have shown that the correlation length is proportional to $W$ (Fig.~\ref{fig:r0_ave_all}A), this is equivalent to plotting $\tau$ vs.\ correlation length apart from an irrelevant numerical factor.  The plot shows that $\tau$ grows with correlation length as expected in a critical system.  Unlike $\xi_o$ vs.\ $W$, the growth of $\tau$ is expected to be a (super-linear) power law, $\tau \sim W^z$, with $z$ called the dynamic critical exponent.  Here we observe a very good power law in some cases (like mice 6 and 7, with $z\approx1.3$), but in other cases the curves deviate downwards from the power law at high $W$. The details of the $\tau$ vs $W$ curves remain to be better understood, in particular given the possibility that the critical power law is altered by arousal changes typical of this experimental model data set, which may be causing a dynamical meandering around the critical point similar to what has been described in earlier experiments \cite{tagliazucchi_criticality_2012, scott2014}.

\begin{figure}
  \includegraphics[width=0.8\columnwidth]{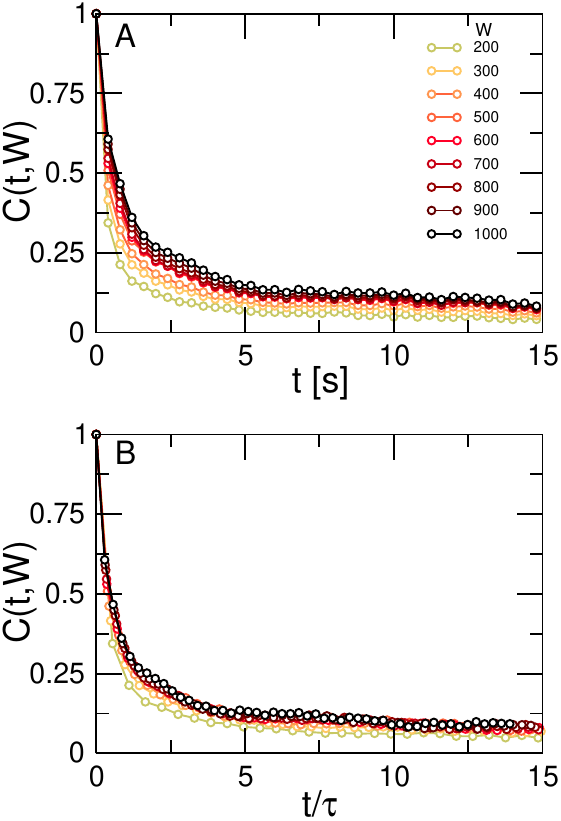}
  \caption{Dynamic scaling of time correlations (A) Normalised time correlation function, $C(t,W)=\CC(t,W)/\CC(0,W)$, see Eq.~\ref{eq:conn-corr-estimate}, vs.\ time for different observation boxes (computed averaging over all boxes and 9 rotations as in Figs.~\ref{fig:methods} and~\ref{fig:tau_example}).  (B) Same data plotted vs.\ the scaling variable $t/\tau$.  The collapse shows that all correlations decay with the same shape, with only the time scale changing for different $W$.  Data are for mouse 7, plane 5.}.
  \label{fig:Ct_scaling}
\end{figure}

Another characteristic of critical dynamics is the scaling of the correlation function itself.  On changing the observation scale (in our case, $W$) one expects that, together with the characteristic time, the shape of the correlation change.  However, \emph{if} the observation scale is changed so that the ratio of the observation scale to the correlation length is fixed, dynamic scaling states that the shape of the (normalised) correlation decay will stay the same, and only the decay scale (i.e.\ $\tau$) should change (in other words, when plotted against $t/\tau$ all correlation functions should look the same).  Since we have argued that, the system being scale free, the effective correlation length is proportional to $W$, the time correlations at different box sizes are effectively computed at fixed observation/correlation scale ratio, and they should scale with $t/\tau$.  The results  in Fig.~\ref{fig:Ct_scaling} show  that this is actually the case, although in some other cases the collapse of the functions is less satisfactory (see Appendices).  

\begin{figure}
    \centering
    \includegraphics[width=0.985\columnwidth]{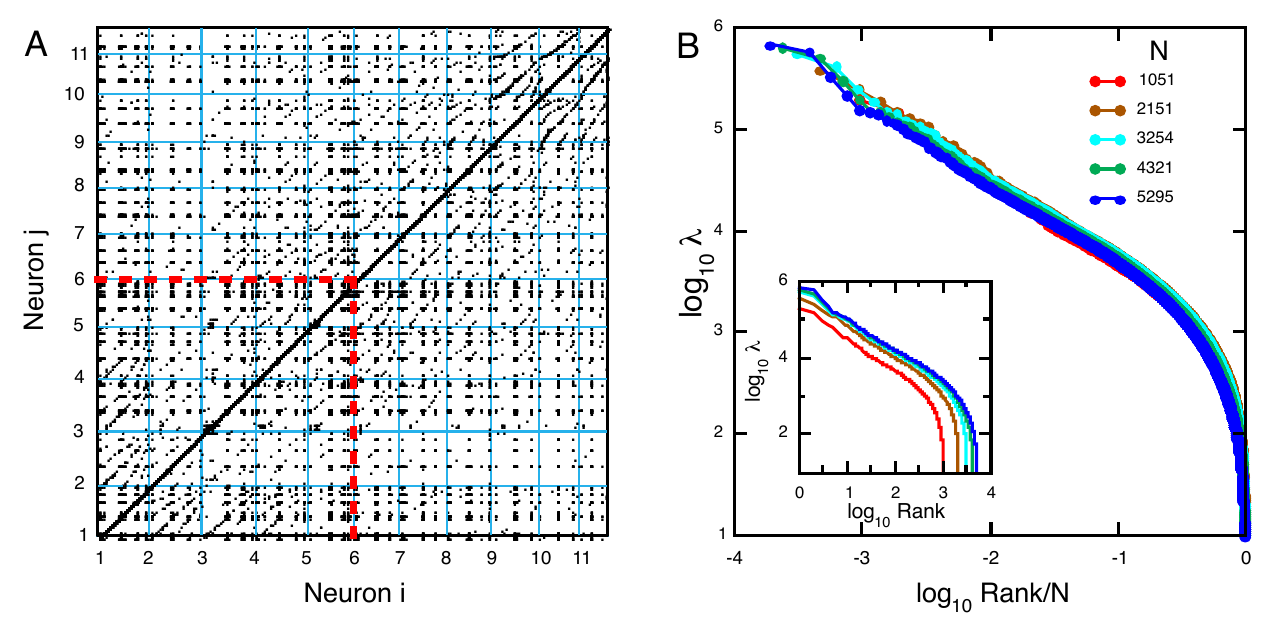}
    \caption{Finite-size scaling of the covariance eigenvalues.  (A)  Schematic view of the covariance matrix of all neurons from mouse 7, including all eleven planes. The black dots denote the Pearson correlation pairs whose values are greater than $1/2$.  The axis labels indicate the planes (from 1 to 11) to which the neurons belong. To help visualise each plane, its boundaries are identified by light blue continuous lines.    (B) Eigenvalues vs.\ rank (double logarithmic) computed from subsets of the covariance matrix in (A).  The  eigenvalues were computed from square subsets of the covariance matrix, of increasing size. To that effect, the covariance data of neurons belonging to the first five consecutive planes  were combined. The largest subset considered (5295 neurons, denoted by the dashed red line in Panel A) is chosen to stay far from the degenerate case where there are more variables that time points, leading to a spurious linear dependence of the set of variables (a total of $10473$ time points are available for this mouse).
Notice that the raw eigenvalue vs.\ rank curves (inset) can be successfully scaled on a single curve by plotting them against the relative rank (rank/matrix size), as shown in the main plot.} 
    \label{fig:eigen}
  \end{figure}

\paragraph{Scale invariance of Eigenvalues---} Systems which, like the present one, exhibit scale-free correlations are expected to show similar invariance also in the eigenvalues of their covariance matrix.  This is worth discussing in this context, because it is frequent in the related literature to find remarks on the fact that the first two or three principal components suffice to explain more than 90\% of the variance .  It may well be that the common explanation behind such observations is a mathematical truism, that necessarily follows from the fact that the system is critical.  Consider Fig.~\ref{fig:eigen}, which depicts the full correlation matrix for all planes of one mouse combined together with the eigenvalues $\lambda_i$ of several subsets of neurons of different sizes.  The eigenvalues are sorted and plotted against their rank in a double logarithmic plot that makes it clear that the magnitude of the eigenvalues decreases as a (negative) power of its rank up to a rank of about half the matrix size.  The sum of the first few terms of a power-law series make up for a sizeable fraction of the total sum, so that a scale-free distribution of the covariance eigenvalues can explain the common observation that the first few principal components explain most of the variance of the cortical population' activity.  
 Interestingly, the $\lambda$ vs.\ rank curves collapse under a single scaling curve when plotted against the relative rank (Fig.~\ref{fig:eigen} panel B).   This finite-size scaling property can be seen as another manifestation of the lack of an intrinsic scale for correlations: the magnitude of the largest eigenvalue is given by the system size.  Also, the other apparent scale, namely the rank at which the power law is cut-offed, is also set by system size.

\section{Conclusions}
\label{sec:conclusions}

We have analysed the space and time correlations of a population of about ten thousand neurons in a region of the mouse visual cortex.  We have observed clear indication that pairs of neurons tend to decorrelate the further apart in space they are.  That the Euclidean distance is a relevant variable affecting the degree of correlation comes as a conclusion of four different ways of measuring correlations in space.  Moreover, we have seen that the characteristic spatial scale of correlation decay scales linearly with the (spatial) observation window $W$.  This is evidence that the correlation decay is scale-free: the only spatial scale is that which is imposed on the system from the outside, i.e.\ the size of the observation box, or eventually the size of the system itself.  A scale-free decay is long range, in the sense that it is described by a power law rather than by an exponential.  So, although correlations do decay with distance, they do so rather slowly.  This new evidence confirms previous studies finding scale-free correlations, and thus suggesting that the resting brain is at or near a critical point.

The main novelty of this study is that we have computed the concomitant time correlations, using the same idea of a varying observation scale.  In this way we have explored how correlations in time and space are related.  We found that the correlation time $\tau$ grows for larger observation scales, much like the spatial correlation scale.  The relation between $\tau$ and $W$ is not linear.  According to dynamic scaling, it is expected to be a power law, which we find for some mice.  The details of this curve and the reason for its departure from a power law in some cases remain to be elucidated, but the important point here is that correlation length and time are closely related, and that the fact that neuron activity is strongly correlated is influencing the collective dynamics, similar to what happens in thermodynamic critical systems.

We have also found that time correlations at different $W$ scale with $t/\tau$, i.e.\ that the decay is identical apart from a time rescaling.  This is in agreement with the expectations of dynamic scaling, but only if the correlation and observation lengths scale together.  This is further evidence for scale-free correlations, because it means that the correlation scale has changed on changing the observation box in absence of any other alteration of the system, which can only happen if the only correlation scale is the observation scale, i.e.\ the system is scale free.

The uncovered  behavior of the time correlations  may be relevant to provide an alternative  mechanistic explanation for the heterogeneity of the  so-called ``temporal receptive fields of integration'' which is established by examining the autocorrelation function of spike-counts at rest \cite{ogawa2010, chen2015a}.  The current interpretation of slow autocorrelation decay in a given neuron is that such neuron is involved in integrating information across long periods of time and viceversa.  This view has been used to support the idea that in the cortex there is a hierarchy of temporal receptive fields \cite{murray2014}, including areas with long decay times which correspond to cognitive tasks requiring long integration of information across time, such as decision making and working memory \cite{cavanagh2016}.  Since dynamic scaling specifically predicts slower decay for larger cortical networks at criticality, it would be interesting to explore if/how this hierarchy corresponds simply to a hierarchy of sizes of the corresponding networks.

Finally, we have shown that finite-size scaling also applies to the eigenvalues of the covariance matrix, a fact that is another manifestation of scale invariance in correlations, and that may explain the frequent observation that  a few principal components account for most of the variance in cortical networks data.  In summary, the combined evidence of spatial correlations, temporal correlations and the eigenvalue analysis builds a stronger case in support of the view that the ongoing brain dynamics is critical or near-critical.

\emph{Acknowledgments:} Supported by Grant No. 1U19NS107464-01 from NIH BRAIN Initiative, by CONICET (Argentina) and Escuela de Ciencia y Tecnología, UNSAM, UNLP (Argentina) and by the Foundation for Polish Science (FNP) project TEAMNET ``Bio-inspired Artificial Neural Networks'' (POIR.04.04.00-00-14DE/18-00). The open-access publication of this article is supported in part  by the program ``Excellence Initiative – Research University'' at the Jagiellonian University in Krakow. Work conducted under the auspice of the Jagiellonian University-UNSAM Cooperation Agreement.

\cleardoublepage

\appendix

\section*{Appendices}

\section{Computation of the correlation time}
\label{sec:correlation-time}

To compute the correlation time we use the definition obtained from $\tilde\rho(\omega)$, the Fourier transform of $\rho(t)=C(t)/C(t=0)$. Normalization of $\rho(t)$ implies that $1=\rho(0) = \int_{-\infty}^{\infty} \frac{d\omega}{2\pi} \,\tilde\rho(\omega)$.  Then a characteristic frequency $\omega_0$ (and a characteristic time $\tau_0=1/\omega_0$) can be defined such that half of the spectrum of $\tilde\rho(\omega)$ is contained in $\omega\in[-\omega_0,\omega_0]$ \cite{halperin_scaling_1969}, i.e.\
\begin{equation}
 \int_{-\omega_0}^{\omega_0} \frac{d\omega}{2\pi} \,\tilde\rho(\omega)=\frac{1}{2}.
 \label{eq:HHomegak}
\end{equation}
This definition of $\tau_0=1/\omega_0$  can be expressed directly in the time domain writing
\begin{equation}
  \begin{split}
    \frac{1}{2} & = \int_{-\omega_0}^{\omega_0} \frac{d\omega}{2\pi}
    \int_{-\infty}^\infty \!\!dt \, \rho(t) e^{i\omega t} =
    2 \int_0^\infty \!\!dt \rho(t) \int_{-\omega_0}^{\omega_0} \frac{d\omega}{2\pi}
    e^{i\omega t} \\
    &= \frac{2}{\pi} \int_0^\infty \!\!dt \, \rho(t)
    \frac{\sin\omega_0 t}{t},
  \end{split}
\end{equation}
where we have used the fact that $\rho(t)$ is even.  Then the correlation time is defined by
\begin{equation}
  \label{eq:HHrelaxtime}
   \int_0^\infty \!\! \frac{dt}{t} \, \rho(t)
   \sin\left(\frac{t}{\tau_0}\right) = \frac{\pi}{4}.
\end{equation}

It can be seen that if $\rho(t) = f(t/\tau)$, then $\tau_0$ is proportional to $\tau$ (it suffices to change the integration variable to $u=t/\tau$ in the integral above). An advantage of this definition is that it copes well with the case when inertial effects are important and manifest in (damped) oscillations of the correlation function

\section{Correlation and distance}
\label{sec:correlation-distance}
\begin{figure}[hb!]
\centering
  \includegraphics[width=.8\columnwidth]{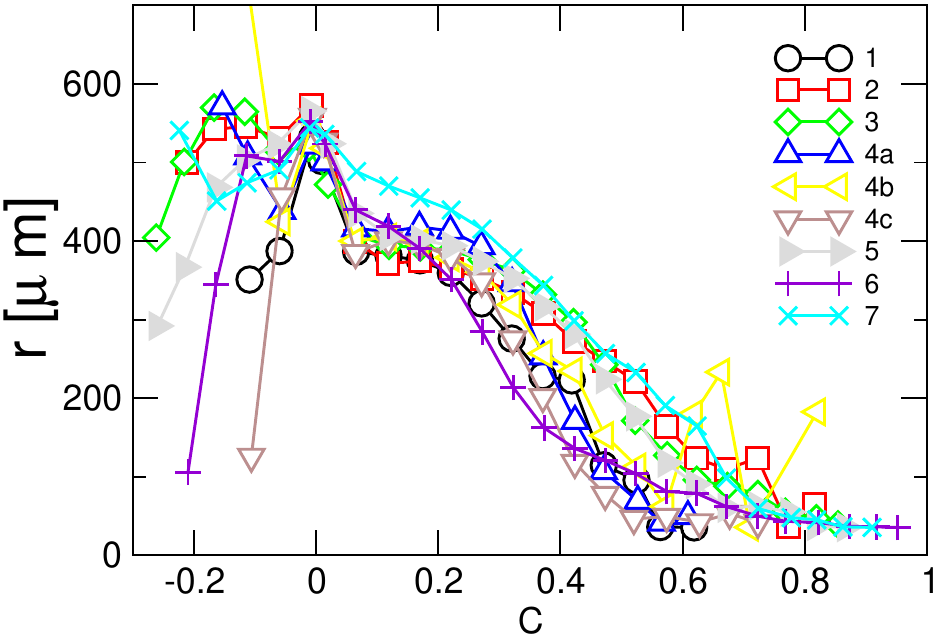}
 \caption{Average pair distance as a function of the Pearson correlation for all mouses and planes.  }
  \label{fig:r_vs_C}
\end{figure}
The correlation functions studied in the main text proceed by choosing a pair of neurons within a given distance and computing their correlation at a single time frame.  This procedure finds that correlations are smaller for larger distances.  An alternative procedure, that leads to the same conclusion, is shown in Fig.~\ref{fig:r_vs_C}.  For each mouse's dataset, one starts by computing the Pearson correlation for all pairs.  After that, correlations are binned (bin width$=.01$).  Finally for the pairs within each bin, their average correlations and their respective average Euclidean distances are computed.

\section{Time correlation function}
\label{sec:time-correlation-SI}

\begin{figure}[hb!]
\centering
  \includegraphics[width=.8\columnwidth]{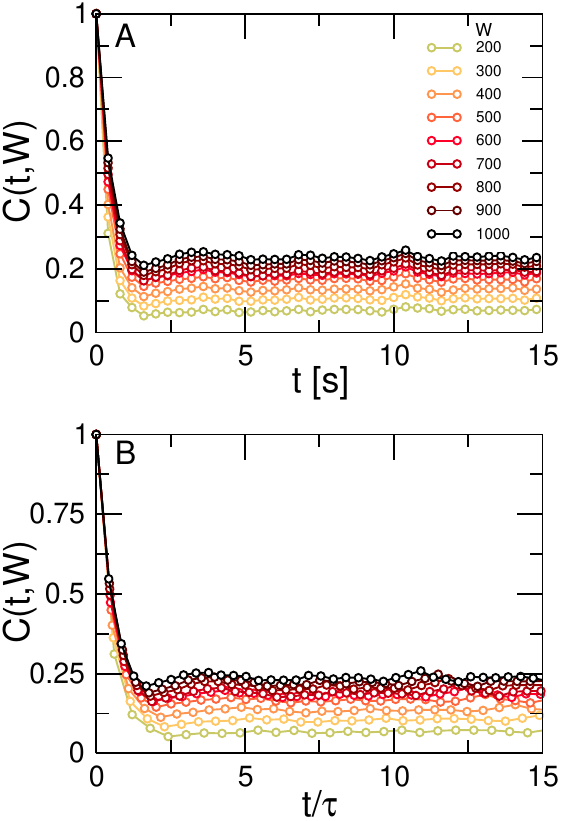}
 \caption{Time correlation function Eq.~\ref{eq:conn-corr-estimate} vs.\ time for different observation boxes. In this example the dynamic scaling of $C(t,W)$ is only apparent for small values of $t$, but not for values larger than 1--2$\,$s. (A) (same format as for Figure \ref{fig:Ct_scaling} in main text) shows the un-normalized $C(t)$ and (B) corresponds to the same data plotted vs.\ the scaling variable $t/\tau$.  Data are for mouse 6, plane 2. }
  \label{fig:r_vs_C_bad}
\end{figure}

In some datasets we noted that $C(t,W)$ only collapses for small values of $t$ but not for longer ones. An example of this disagreement is presented in Fig.~\ref{fig:r_vs_C_bad}. With the present data, we can only provide probable reasons.  The first is related with  non-stationarity conditions linked with the fact that the data is obtained while the animal executes at will bursts of  wheel running.  This alone may affect the entire correlation structure of the brain.  The second factor may simply be the changes in arousal, which in these experiments was monitored by changes in the mice pupil diameter. The role both possibilities deserves to be explored in further work.

\section{Null hypotheses and preprocessing artifacts}
\label{sec:artifacts} 

\subsection{Insights from numerical simulations}

To test for potential artifacts in the results, we have conducted a series of numerical simulations for the GH \cite{GH} model simulated as in  Aguilar Trejo \emph{et al.} \cite{Eyisto}.  In brief, we simulate a square lattice of $L\times L$  (here $L=600$) neurons with periodic boundary conditions.  Each neuron is first connected to the closest  $k = 24$ neighbors, and then each output connection  is rewired with probability $\pi = 0.01$  to another neuron within the whole system. Nonzero connection weights $W$ are taken from a random  distribution $p(W = w)\propto exp(-w/\lambda)$  with $\lambda =12.5$. The connection matrix is fixed throughout the simulations.  Each neuron  $i$ is a cellular automaton that can be in any of 3 states, as a function of time: quiescent ($S_i(t) = 0$), active ($S_i(t) = 1$) or refractory ($S_i(t) = 2$).  A quiescent neuron can become active in the next time either by spontaneous activation (with probability  $r_1=10^{-5}$) or if  the contribution of its active connections at time $t$ is larger than the threshold $T$ (i.e., if $\sum_j W_{ij} \delta_{S_j(t),1}>T$). An active neuron will always became refractory, and a refractory neuron will become quiescent with probability $r_2=0.3$.  This model udergoes a continous phase transition as $T$ is varied: it can be found in the  supercritical (very active and bursting, for $T<T_C$), critical (for $T= T_C \simeq 0.318$) and subcritical (for $T>T_C$) states.

\subsection{Slow decay of Ca$^{2+}$ does not affect the correlation length scaling results}
\label{sec:slow-decay-ca2+}

\begin{figure}
\centering
\includegraphics[width=\columnwidth]{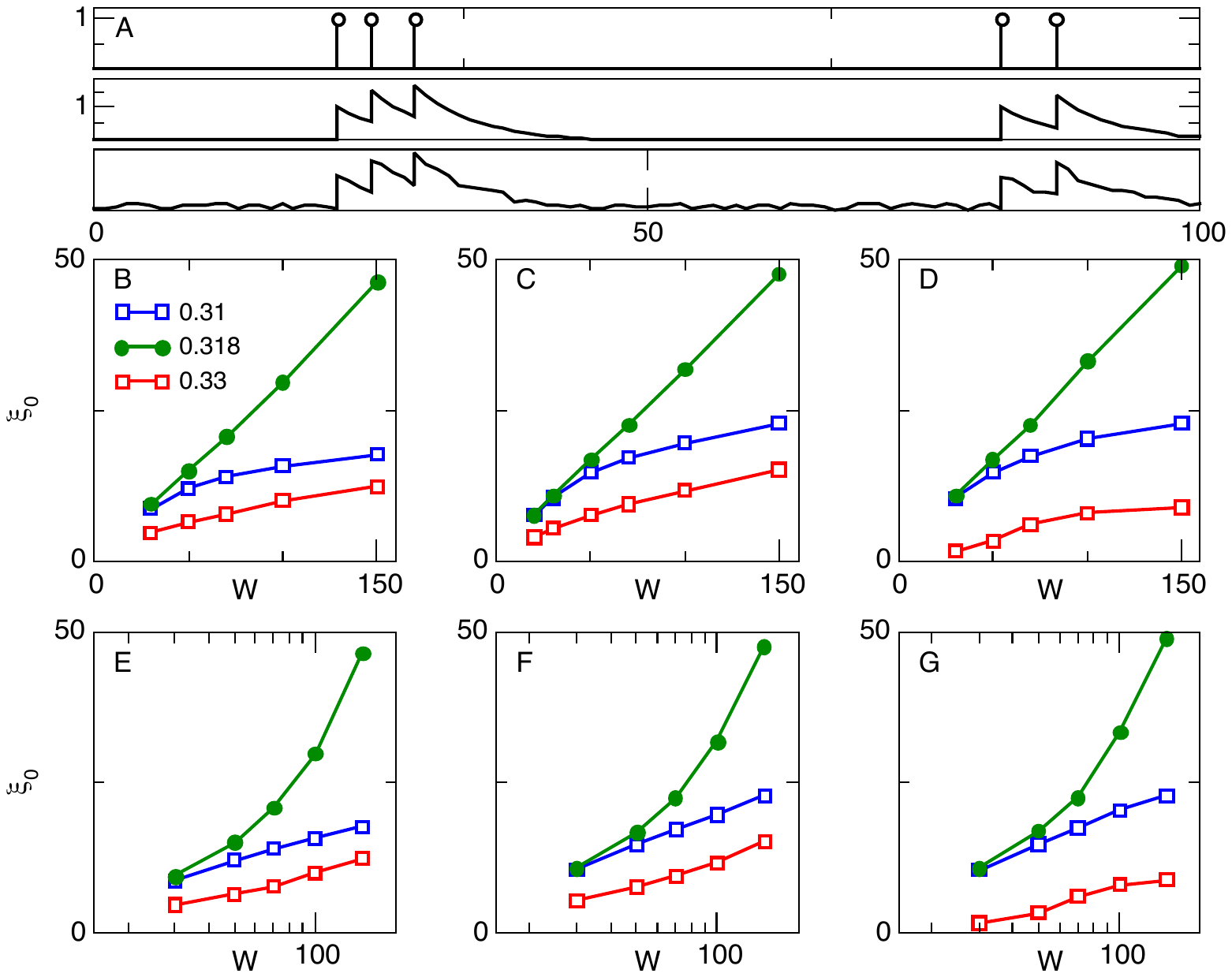} 
\caption{A slow decay does not affect the scaling of the  correlation length by using the CCF.  (A) example of the  signal used to compute $C(r)$ for a single random neuron at $T=T_C$. Top: spike time signal, middle: spikes convolved with an exponentially decaying function with characteristic time $\tau_0=5$, bottom: convolved spikes plus a random noise of amplitude 0.2;   (B) Correlation length $\xi_0$ as a function of window size for subcritical ($T=0.33$), critical ($T=0.318$) and supercritical ($T=0.31$) states, using the spike time signal. (C) Same as panel B, using the convolved signal.  (D) Same as panel B for the noisy convolved signal. (E), (F)  and (G) same as Panels B, C and D in log-linear axis.  Simulations used a lattice of $600\times600$ sites with periodic boundary conditions, recorded for 30000 time steps. All other parameters as in ref.~\cite{Eyisto}.  \label{convolved}} 
\end{figure}

To exclude possible effects of the slow temporal decay of the calcium signal on the scaling of correlations reported here,  we compare the CCF from both a discrete time-series of spikes and from a ``Calcium-like'' exponentially decaying function of time triggered by each spike.  The Calcium-like time-series is built by convolving the spike time-series with an exponential with characteristic time $\tau_e=5$.  An example of the original and the convolved signals are shown in Fig.~\ref{convolved}-A, while  $\xi_0$ vs.\ $W$  can be found in panels B, C, E and F.  It can be seen that the scaling of $\xi_0$ with window size $W$ is the same for the spike-like and the Calcium-like time-series: for the critical state, $\xi_0$ grows linearly with $W$, while the growth is logarithmic for sub/supercritical states.  This result is not surprising since we are computing  how correlations of the fluctuations around the mean scale with distance, and not simply the correlation of $S(t)$ between two places. To further illustrate the point we recomputed the results after adding noise (with uniform distribution of width 0.2) to each neuron at each time step (signal to noise ratio $\simeq 2$).  Note that despite the noise, the scaling of $\xi_0$ is still unchanged (see Fig.~\ref{convolved}-D and G).
 
\subsection{Slow sampling rate does not affect the scaling of correlation length}
  
\begin{figure}
\centering
\includegraphics[width=0.9\linewidth]{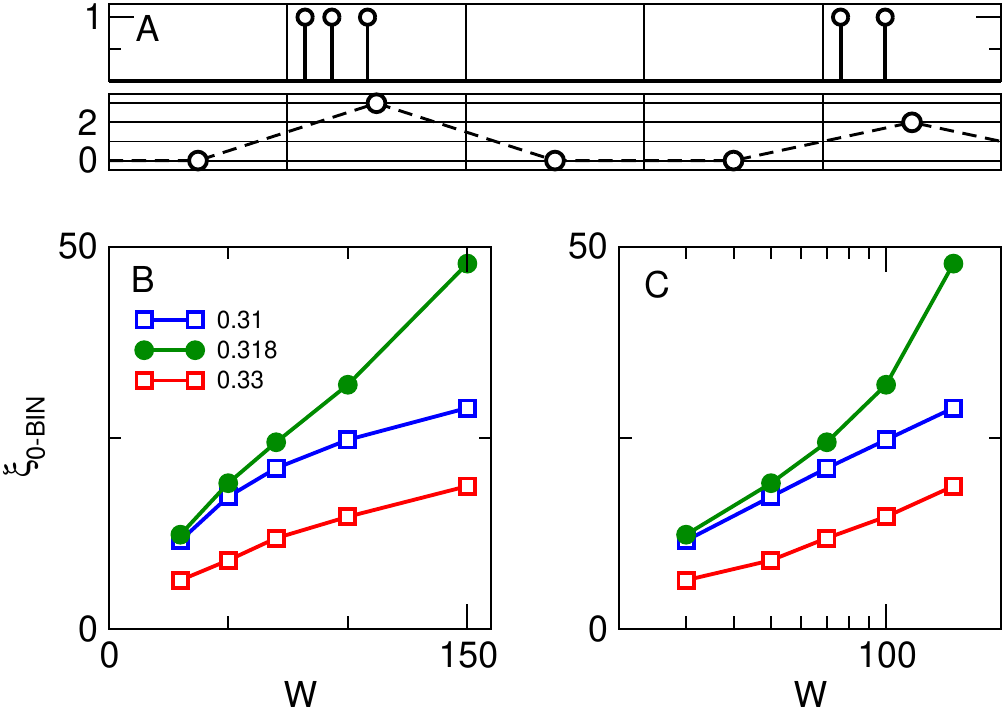} 
\caption{Time binning does not affect the scaling of $\xi_0$. 
(A) example of the spike signal for a single randomly chosen neuron at $T=T_C$. Top: spike time signal; bottom: spikes per frame (dashed lines are a guide to the eye).  (B) Correlation length $\xi_{0-BIN}$  vs.\  window size for subcritical ($T=0.33$), critical ($T=0.318$) and supercritical ($T=0.31$) states, computed for the spikes-per-frame data, (C) Same as panel B in log-linear axis. Results computed for 200000 time steps, binned into 10000 bins of 20 steps each. All other parameters as in Fig.~\ref{convolved}.  \label{FigTimeBinning}} 
\end{figure}

Next, we consider the potential effects of the relatively slow sampling rate in the fluorescence recording (3$\,$Hz). To mimic the experimental protocol, we run numerical simulations where the signal is time binned: the sum of the activity over 20 frames (instead of the instantaneous activity) is used as signal to compute $C(r)$. The results for this Binned correlation length, termed $\xi_{0-BIN}$ are shown in Fig. \ref{FigTimeBinning}.  We conclude that time binning does not alter the scaling.

\subsection{Common drives do not affect the scaling of correlation length}
  
\begin{figure}
\centering
\includegraphics[width=0.9\linewidth]{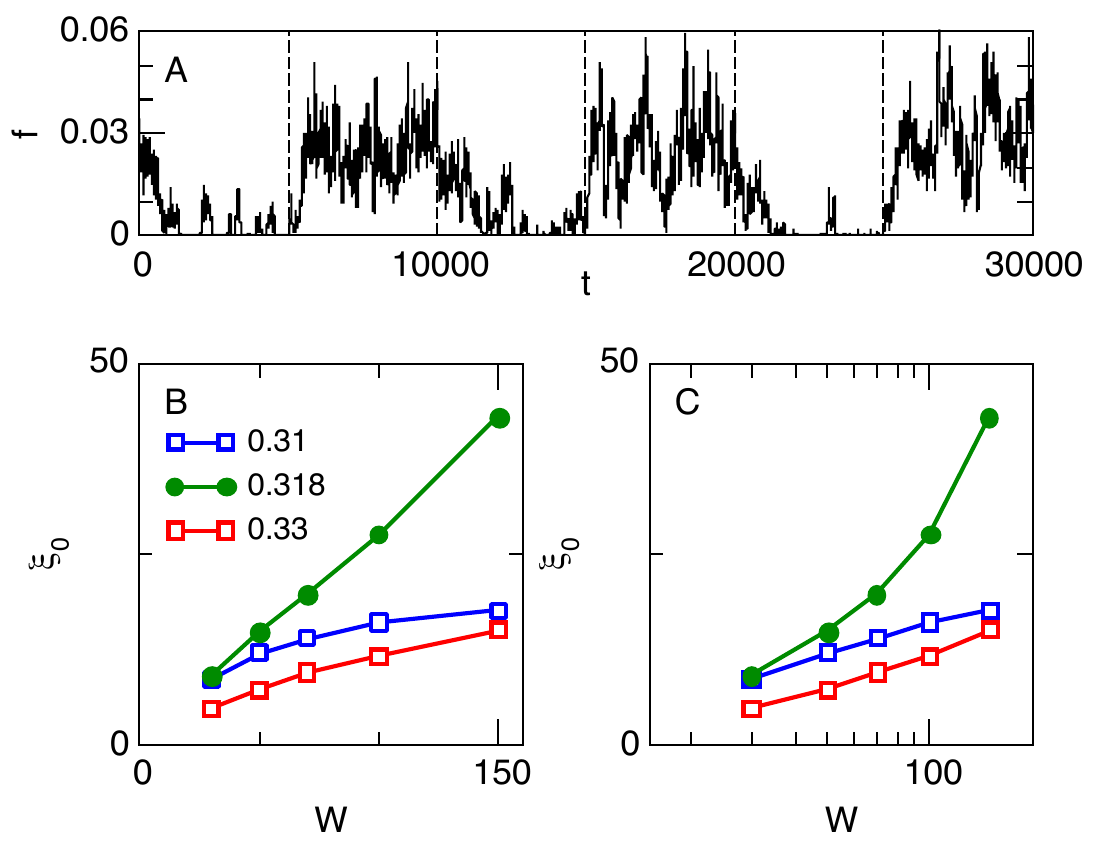} 
\caption{Arousal-like changes in the spontaneous activity do not affect the correlation scaling behavior. (A) fraction of active neurons ($f$) as a function of time for 30000 time steps, where the value of the spontaneous activation probability $r_1$ is changed every 5000 time steps  between a low ($r_1=10^{-5}$) and a high  ($r_1=2 \cdot 10^{-4}$) value. (B) and (C) show the correlation length as a function of window size, in linear and log-linear  axes, for subcritical ($T=0.33$), critical ($T=0.318$), and supercritical states ($T=0.31)$. All other parameters as in Fig.~\ref{convolved}. }\label{Arousal}
\end{figure}

To show that the connected correlation function is insensitive to common drives,  such as slow network-wide arousal changes,  we conducted numerical simulations where the spontaneous activation rate $r_1$ is periodically changed between a low and a high value ($r_1=10^{-5}$ and $r_1= 2 \cdot 10^{-4}$ respectively) every 5000 time steps (Fig.~\ref{Arousal}).  Note that these variations in the spiking rate are not changing the dynamical state of the network (i.e., we keep $T=T_C$).  This result demonstrates that even large changes in the rate are not preventing the observation of the scale free correlation behavior since, as already discussed, the CCF subtracts the population mean instantaneous activity (Eq.~\ref{eq:connected_corr}).

\subsection{The scaling of correlation time is not due to Ca$^{2+}$ slow decay}
 
\begin{figure}
\centering
\vspace{0.5cm}
\includegraphics[width=0.95\linewidth]{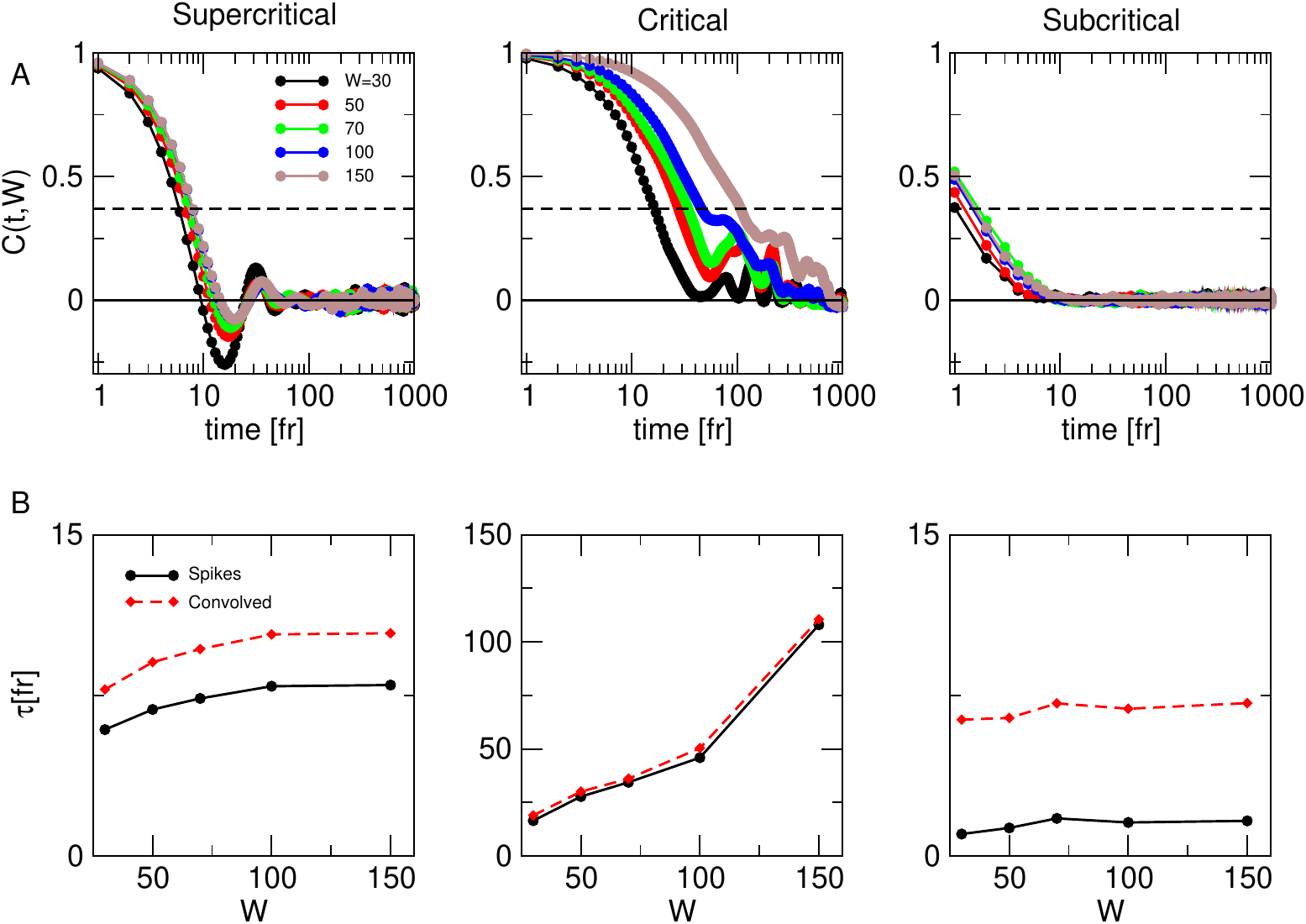} 
\caption{
Connected correlation time in the numerical model is not originated by slow (Calcium-like) decay. (A) Results computed from the spike data for the supercritical ($T=0.31$, left), critical ($T=0.318$, middle) and subcritical ($T=0.33$, right) state.  (B) Correlation time $\tau$ estimates, computed  as the crossing of $C(t,W)=1/e$, using the spike data of panel A (circles) and the Calcium-like convolved data already described in Fig. \ref{convolved} (squares). Note that at criticality both estimations of $\tau$ increase with window size $W$  but remains constant for sub or supercritical dynamics.}\label{FigTauSim}
 \end{figure}

We can safely exclude the possibility that the dynamic scaling is an artifact of the slow decay of the  Ca$^{2+}$ signals from which the authors of ref.~\cite{stringer2019} extracted the time series. This is demonstrated by the results in Fig.~\ref{FigTauSim} which show the time correlation function and the correlation times as a function of window size, for the spike-like time series and the Calcium-like time series considered in Sec.~\ref{sec:slow-decay-ca2+} above.  Here we estimate the correlation time as the crossing of the value $1/e$ of the connected time correlation function $C(t,W)$, for the spike and the Calcium-like convolved data.  Numerical results show as expected that the correlation time is small and almost independent of window size for subcritical ($T=0.33$) and supercritical ($T=0.31$) states, while  the critical state shows a  much larger correlation time, strongly dependent on window size.  More importantly, note that the correlation times estimated from the spike-like and from the convolved Calcium-like data only differ by a constant for all regimes, showing the same qualitative behavior as a function of $W$.

\subsection{The scaling of correlation length is destroyed by random position permutations}

In order to check for potential artifacts affecting the overall network, we tested how the correlations from the mice and the simulation data change when the neuron positions are randomly shuffled. For that we re-assigned the timeseries of each neuron to another, randomly chosen neuron.  Shuffling the data in this way should not change results if the  correlations were due to an external or hidden variable driving the entire network activity.  
  
\begin{figure}
\centering
\includegraphics[width=\columnwidth]{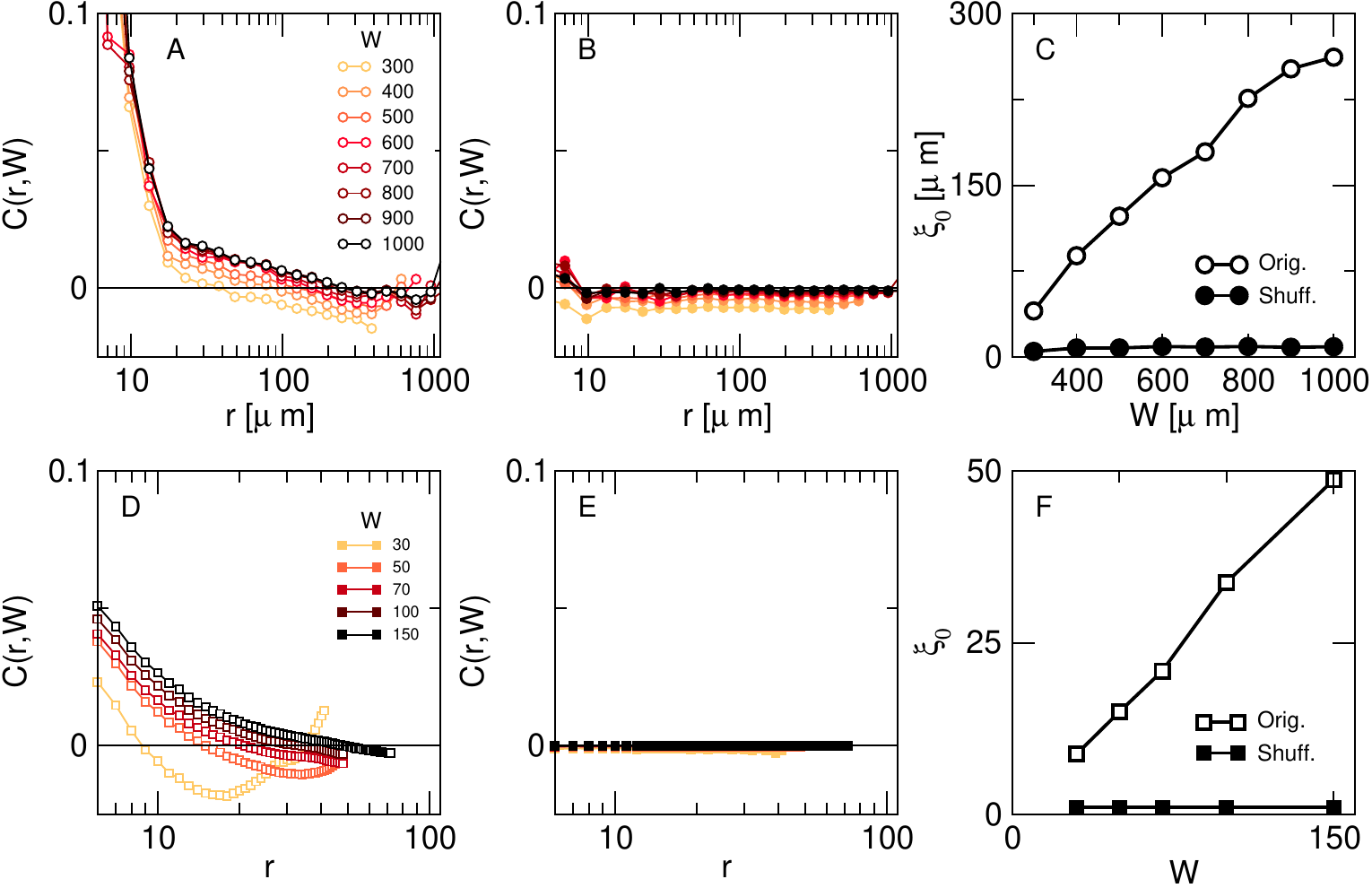} 
\caption{The scaling of correlation length is destroyed by random position permutations. Top panels (A,B and C) correspond to Mouse 4c, layer 4 data (using a single window for each value of W , without rotations) and bottom panels (D,E and F) to numerical simulations of the model at the critical regime (i.e.$, T= T_C$). 
The plots on the left column depict the CCF computed from the original timeseries (for experiments in A and for simulations in D) while the plots on the middle column to the CCF computed  after shuffling randomly the neuron positions. Panels C and F shows the correlations lengths as function of the window size $W$ extracted from all the conditions.   
 All  other model parameters as in Fig.~\ref{convolved}.} \label{FigNull}
\end{figure}

In Fig.~\ref{FigNull} we show  how the correlations are changed by this shuffling,  using a single window (without rotations) of mouse 4c, layer 4. As expected from the theory, the correlation lengths computed from the shuffled data are very small and \emph{finite}, i.e., $\xi_0$ is independent of $W$. Similar results hold for all windows of all layers and all mice. The numerical simulation results shown in the bottom panels of the same figure demonstrate the same behavior when the data gathered from the model at the critical state is shuffled in the same way.
These results exclude the possibility of attributing the origin of the scaling of correlation length to unknown hidden variables driving the entire network correlations and support the argument for critical dynamics.

\end{document}